\documentclass[aps,prb,showpacs, letterpaper,twocolumn,superscriptaddress]{revtex4}

\usepackage{amssymb}
\usepackage{amsmath}
\usepackage{latexsym}
\usepackage[pdftex]{graphicx}
\usepackage{placeins}
\usepackage{booktabs}
\usepackage{epstopdf}
\usepackage{subfig}

\newcommand{\ie}{\textit{i.e.}{}\ }
\newcommand{\eg}{\textit{e.g.}{}\ }

\newcommand{\ud}{\,\mathrm{d}}

\newcommand{\mc}[1]{\mathcal{#1}}

\newcommand{\exc}{\epsilon_{\mathrm{xc}}}
\newcommand{\ext}{\mathrm{ext}}
\newcommand{\eff}{\mathrm{eff}}

\newcommand{\jump}[1]{[[#1]]}
\newcommand{\mean}[1]{\{\{#1\}\}}

\renewcommand{\H}{\mathsf{H}}
\newcommand{\M}{\mathsf{M}}
\newcommand{\W}{\mathsf{W}}
\renewcommand{\L}{\mathsf{L}}
\renewcommand{\S}{\mathsf{S}}
\newcommand{\C}{\mathsf{C}}
\newcommand{\V}{\mathsf{V}}
\newcommand{\G}{\mathsf{G}}
\newcommand{\I}{\mathcal{I}}

\newcommand{\tot}{\mathrm{tot}}

\newcommand{\abs}[1]{\lvert#1\rvert}

\newcommand{\average}[1]{\left\langle#1\right\rangle}

\providecommand{\abs}[1]{\lvert#1\rvert}

\providecommand{\average}[1]{\left\langle#1\right\rangle}

\providecommand{\bra}[1]{\langle#1\rvert}
\providecommand{\ket}[1]{\lvert#1\rangle}

\def\au{\text{a.u.}}

\begin{document}

\title{Element orbitals for Kohn-Sham density functional theory}
\affiliation{Computational Research Division, Lawrence Berkeley National
Laboratory, Berkeley, CA 94720, USA} 
\affiliation{Department of Mathematics and ICES, University of Texas at
  Austin, Austin, TX 78712, USA}
\author{Lin Lin}
\email{linlin@lbl.gov}
\affiliation{Computational Research Division, Lawrence Berkeley National
Laboratory, Berkeley, CA 94720, USA} 
\author{Lexing Ying}
\affiliation{Department of Mathematics and ICES, University of Texas at
  Austin, Austin, TX 78712, USA}

\pacs{71.15.Ap, 71.15.Nc}

\begin{abstract}
  We present a method to discretize the Kohn-Sham Hamiltonian matrix in
  the pseudopotential framework by a small set of basis functions
  automatically contracted from a uniform basis set such as planewaves.
  Each basis function is localized around an element, which is a small
  part of the global domain containing multiple atoms.  We demonstrate
  that the resulting basis set achieves meV accuracy for 3D densely
  packed systems with a small number of basis functions per atom. The
  procedure is applicable to insulating and metallic systems.
\end{abstract}

\maketitle

\section{Introduction}\label{sec:intro}

Kohn-Sham density functional theory (KSDFT)~\cite{KohnSham:65} is the
most widely used electronic structure theory for condensed matter
systems.  When solving the Kohn-Sham equations, the choice of basis
functions usually poses a dilemma for practitioners. The accurate and
systematically improvable basis functions that are uniform in space,
such as plane waves or finite elements, typically result in a large
number of degrees of freedom ($500\sim 10000$) per atom in the framework
of norm conserving pseudopotential~\cite{TroullierMartins1991}
especially for transition metal elements.  The number of basis functions
per atom can be reduced to the order of hundreds using ultrasoft
pseudopotential~\cite{Vanderbilt1990}, or augmentation techniques in the
core-region such as the linearized augmented plane-wave (LAPW)
method~\cite{Andersen1975} and the projector augmented wave (PAW)
method~\cite{Blochl1994}.  The relatively large number of basis functions used
leads to a large prefactor in front of the already expensive cubic
scaling for solving KSDFT.  

Contracted basis functions, such as Gaussian type orbitals, atomic
orbitals or muffin-tin orbitals, can represent the Kohn-Sham orbitals
with a small number of degrees of freedom per atom ($4\sim 100$). These
contracted basis functions contain a number of parameters to be
determined.  The flexibility for choosing different forms of parameters
has generated a vast amount of
literature (see \eg
Refs.~\onlinecite{Ozaki:03,BlumGehrkeHankeEtAl2009,Junquera:01,ChenGuoHe2010,AndersenSaha2000,QianLiQiEtAl2008})
in the past few decades, which has been reviewed recently in
Ref.~\onlinecite{BowlerMiyazaki2012}.  The parameters in the contracted
basis functions are typically constructed by a fitting procedure for a
range of reference systems.  The necessary inclusion of polarization
basis~\cite{FrischPopleBinkley1984}, diffuse basis
~\cite{ClarkChandrasekharSpitznagelEtAl1983}, multiple radial functions
for each angular momentum (multiple $\zeta$ basis~\cite{Junquera:01})
are just a few examples when the choice of basis functions becomes
difficult and system dependent especially for complex systems. 

It is desirable to combine the advantage of uniform basis set in which
the accuracy is controlled by no more than a handful of universal
parameters for almost all materials, and the advantage of contracted
basis functions with a very small number of basis functions per atom.
In other words, we would like to generate a small number of contracted basis functions
by a unified procedure with high accuracy comparable to that obtained
from uniform basis functions.  In a recent work~\cite{LinLuYingE2012},
we have developed a unified method for constructing a set of contracted
basis functions from a uniform basis set such as planewaves in the
pseudopotential framework. The new
basis set, called adaptive local basis (ALB) set, is constructed by
solving the Kohn-Sham problem restricted to a small part of the domain
called {\em element}.  Each ALB is discontinuous from the perspective of
the global domain, and the continuous Kohn-Sham orbitals are
approximated by the discontinuous ALBs under a discontinuous Galerkin
framework~\cite{Arnold:82}.  It was demonstrated that the ALBs are able
to achieve high accuracy (in the order of $1$ meV) using disordered Na and
Si as examples.  However, the number of basis functions per atom
increases with respect to dimensionality.  For example, $40$ basis
functions per atom is needed to reach the accuracy of $1$ meV/atom for 3D
bulk Na system.  

In this paper, we propose a new basis set that is constructed from
linear combination of adaptive local basis functions.  Each new basis
function, dubbed {\em element orbital} (EO), has localized nature
around its associated element of the domain.  The number of EOs used is
significantly reduced compared to the number of ALBs for 3D bulk
systems.  We demonstrate that $4$ EOs per atom are
sufficient to achieve $1$ meV
per atom accuracy for 3D bulk Na system with disorderedness.  We also
apply EOs to study Na, Si and graphene, with varying
system sizes, lattice constants or types of defects. The new method
consistently achieves meV accuracy for calculating the total energy
when compared to standard electronic structure software such as
ABINIT~\cite{Abinit1}.  Since the EOs are contracted from
a uniform basis set such as planewave basis set, the shape of the
EOs has more flexibility to reflect the environmental
effect than contracted basis sets which are centered around atoms.
Numerical result indicates that the shape of EOs can
resemble both atomic orbitals of different angular momentum and
chemical bonds centered in the interstitial region, depending on their
chemical environment.

We remark that the construction of the EOs is closely
related to several existing techniques for reducing the number of basis
functions per atom, starting from a large primitive basis set consisting
of Gaussian orbitals or atomic
orbitals~\cite{Ozaki:03,BlumGehrkeHankeEtAl2009,RaysonBriddon2009,Rayson2010}.
However, the EOs are contracted from a fine uniform basis
set such as planewaves, and a number of difficulties arise which
makes the previous techniques difficult to be directly applied.  
For instance, the filtration technique in
Ref.~\onlinecite{RaysonBriddon2009,Rayson2010} constructs a near-minimal basis set
from a large number of Gaussian type orbitals by applying a filtration
matrix to a set of trial orbitals, taken from one or a few Gaussian-type
orbitals.  When the Gaussian-type orbitals are replaced by a fine
uniform basis set such as planewaves, finding a good set of trial
orbitals itself becomes a difficult task, and the construction of trial
orbitals can inevitably introduce a set of undetermined parameters,
which is not desirable in the current framework. 

This paper is organized as follows: Section~\ref{sec:ALB} introduces the
adaptive local absis functions in the discontinuous Galerkin framework
for solving Kohn-Sham density functional theory in the pseudopotential
framework.  The construction of
the element orbitals is introduced in Section~\ref{sec:EO}.
Section~\ref{sec:parallel} discusses briefly the implementation
procedure of element orbitals.  The performance of element orbitals is
reported in Section~\ref{sec:num}, followed by the discussion and
conclusion in Section~\ref{sec:conclusion}.

\section{Adaptive local basis functions}\label{sec:ALB}

Consider a quantum system with $N$ electrons under external potential
by $V_{\ext}$ in a rectangular domain $\Omega$ with periodic boundary
condition. To simplify the equations, we ignore the electron spin for
now. In Kohn-Sham density functional theory at a finite $T=1/(k_B
\beta)$~\cite{KohnSham:65,Mermin1965}, the Helmholtz free energy is
given by
\begin{multline}\label{eq:Merminfunc}
  \mc{F}_{\tot} = 
  \mc{F}_{\tot}(\{\psi_i\}, \{f_i\}) = \frac{1}{2}
  \sum_{i} f_i \int \abs{\nabla
    \psi_i(x)}^2 \ud x \\
  + \int V_{\ext}(x) \rho(x) \ud x   + \sum_{\ell} \gamma_{\ell}
  \sum_{i} f_i \abs{\int b_{\ell}^{\ast}(x) \psi_i(x) \ud x}^2 \\
  + \frac{1}{2} \iint \frac{\rho(x) \rho(y)}{\abs{x-y}} \ud x \ud y   + \int\exc[\rho(x)] \ud x \\
  + \beta^{-1} \sum_i \bigl( f_i \ln f_i +  (1 - f_i) \ln(1 - f_i) \bigr),
\end{multline}
Correspondingly $\{\psi_{i}(x)\}$ and $\{f_i\}$ are the
solutions to the minimization problem
\begin{equation}\label{eqn:KSfunc_finiteT}
  \begin{split}
    &\min_{\{\psi_{i}\},\{f_i\}} \mc{F}_{\tot}(\{\psi_i\},\{f_i\}),\\
    &\text{s.t.} \quad \int \psi_i^{\ast}(x) \psi_j(x) \ud x =
    \delta_{ij},\quad i,j=1,\cdots,\widetilde{N}.
  \end{split}
\end{equation}
$\{f_i\} \in [0, 1]$ are the occupation numbers which add up to the
total number of electrons $N = \sum_{i=1}^{\widetilde{N}} f_i$.  Here we use
exchange-correlation functional under local density approximation
(LDA)~\cite{CeperleyAlder1980,PerdewZunger1981} and adopt norm
conserving pseudopotential~\cite{TroullierMartins1991}, with the
projection vector of the nonlocal
pseudopotential in the Kleinman-Bylander form~\cite{KleinmanBylander:82}
denoted by $\{ b_{\ell}(x) \}$, and $\gamma_{\ell}=\pm 1$ is a sign. The
number of eigenstates $\widetilde{N}$ calculated in practice is chosen
to be slightly larger than the number of electrons $N$ in order
to compensate for the finite temperature effect, following the
criterion that the occupation number $f_{\widetilde{N}}$ is
sufficiently small (less than $10^{-8})$. The electron density is
given by
\[
\rho(x)=\sum_{i=1}^{\widetilde{N}} f_i \abs{\psi_i(x)}^2.
\]

The Kohn-Sham equation, or the Euler-Lagrange equation associated
with \eqref{eqn:KSfunc_finiteT} is~\cite{KohnSham:65,Mermin1965}
\begin{equation}\label{eqn:KS}
  H[\rho] \psi_i = \Bigl( - \tfrac{1}{2} \Delta + V_{\eff}[\rho]
  + \sum_{\ell} \gamma_{\ell} \ket{b_{\ell}} \bra{b_{\ell}} \Bigr) \psi_i 
  = \lambda_i \psi_i,
\end{equation}
where the effective one-body potential $V_{\eff}[\rho]$ is
\[
V_{\eff}[\rho](x) = V_{\ext}(x) + \int \frac{\rho(y)}{\abs{x - y}}\ud y
+ \exc'[\rho(x)]
\]
and the occupation numbers $\{f_i\}_{i\ge 1}$ follow the Fermi-Dirac
distribution 
\[
f_i = \frac{1}{1 + \exp( \beta(\lambda_i - \mu))}.
\]
Here the chemical potential $\mu$ is chosen so that
$\sum_{i=1}^{\widetilde{N}} f_i
= N$. In each SCF iteration of \eqref{eqn:KS}, we freeze $\rho$
and solve for the $\tilde{N}$ lowest eigenfunctions
$\{\psi_i(x)\}_{1\le i \le \tilde{N}}$. This linear eigenvalue problem
is the focus of the following discussion.



The discontinuous Galerkin (DG) framework~\cite{Arnold:82} provides
flexibility in choosing appropriate basis functions to discretize the
Kohn-Sham Hamiltonian $H[\rho]$. In the DG framework, a smooth
function delocalized across the global domain can be systematically
approximated by a set of discontinuous functions that are localized in
the real space. Let $\mc{T} = \{E_1, E_2, \cdots, E_M \}$ be a
collection of {\em elements}, \ie disjoint rectangular partitions of
$\Omega$, and $\mc{S}$ be the collection of surfaces $\{\partial E_{k}\}$
that correspond to each element $E_{k}$ in $\mc{T}$. We associate with
each $E_k$ a set of orthogonal basis functions $\{u_{k, j}(x)\}_{1\le
  j \le J_k}$ supported in $E_k$, with the total number of basis
functions given by
\[
N^b=\sum_{k=1}^{M} J_k.
\] 
Under such a basis set, the Hamiltonian is discretized into an
$N^b\times N^b$ matrix with entries given by
\begin{equation}
  \begin{split}
  &\H (k',j';k,j) \\
  =& 
  \frac{1}{2}\average{\nabla u_{k',j'}, \nabla    u_{k,j}}_{\mc{T}} 
  -\frac{1}{2}\average{\jump{ u_{k',j'}}, \mean{\nabla
  u_{k,j}}}_{\mc{S}} \\
  &-\frac{1}{2}\average{\mean{\nabla u_{k',j'}},    \jump{u_{k,j}}}_{\mc{S}} 
  + \alpha \average{\jump{u_{k',j'}}, \jump{u_{k,j}}}_{\mc{S}} \\
  &+\average{u_{k',j'}, V_{\eff} u_{k,j}}_{\mc{T}} 
  + \sum_\ell  \gamma_{\ell} \average{u_{k',j'},b_\ell}_{\mc{T}} \average{b_\ell, u_{k,j}}_{\mc{T}},
  \end{split}
  \label{eqn:KSDG}
\end{equation}
where $\average{\cdot, \cdot}_{\mc{T}}$ and $\average{\cdot,
  \cdot}_{\mc{S}}$ are inner products in the bulk and on the surface
respectively, and $\alpha>0$ is a fixed parameter for penalizing
cross-element discontinuity. The notations $\mean{\cdot}$ and
$\jump{\cdot}$ stand for the average and jump operators across
surfaces~\cite{Arnold:82}.  Comparing \eqref{eqn:KS} and
\eqref{eqn:KSDG}, the new terms involving the average and jump
operators can be derived from integration by parts of the Laplacian
operator, and provide consistency and stability of the DG
method~\cite{Arnold:02}.

\begin{figure}[h]
  \begin{center}
    {\includegraphics[width=0.40\textwidth]{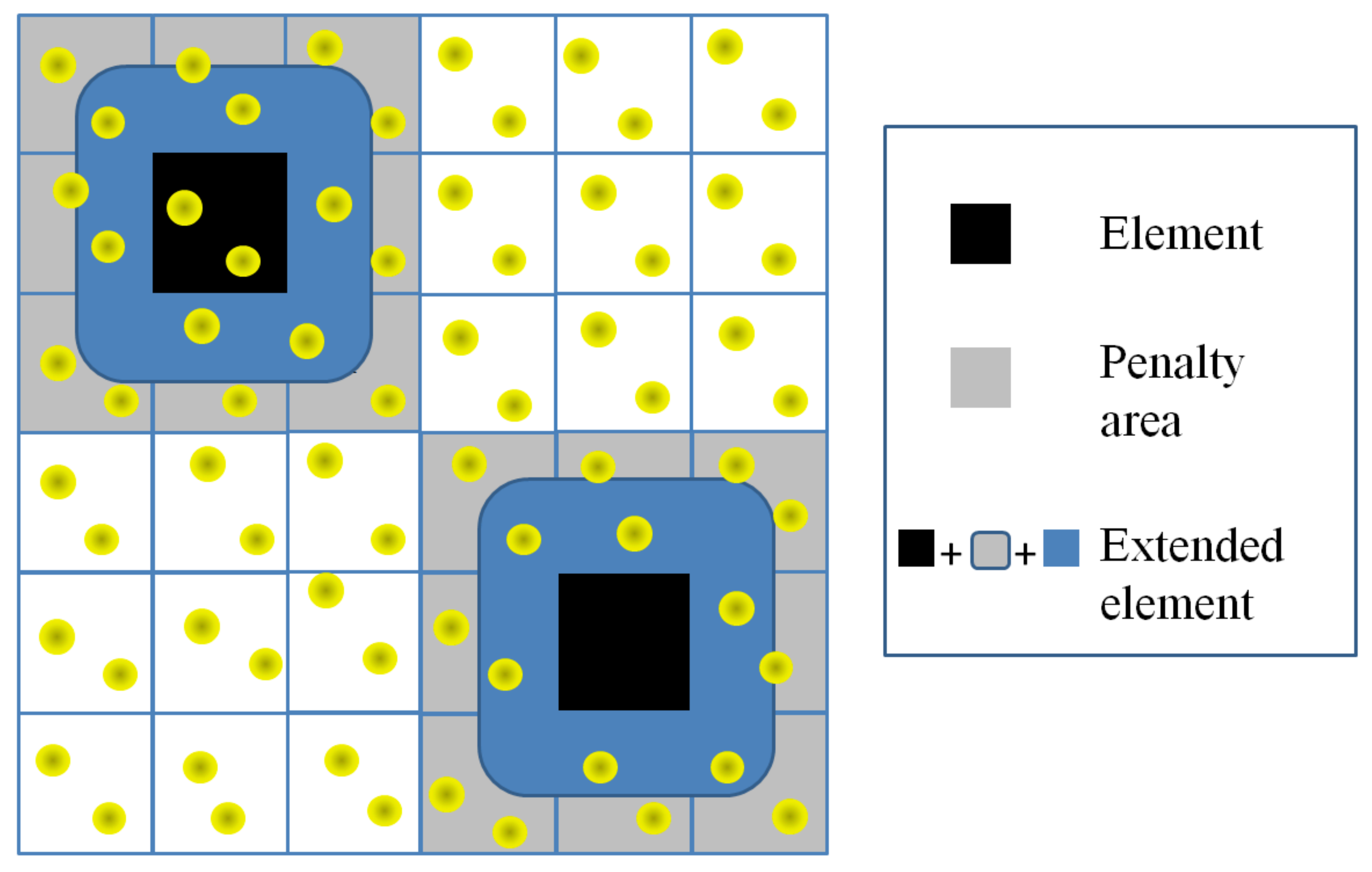}}
  \end{center}
  \caption{(color online) Sketch for the construction of adaptive local
  basis functions and element orbitals.  Each
  adaptive local basis function is supported in an element.  Each
  element orbital is supported in an extended element.}
  \label{fig:cartoon}
\end{figure}

In the work of adaptive local basis set~\cite{LinLuYingE2012}, the
functions $\{u_{k, j}\}_{1\le j \le J_k}$ in each element $E_{k}$ are
determined as follows. Let $d$ be the dimension of the system. For
each $E_k$ (one black box in Fig.~\ref{fig:cartoon}), we define an
associated extended element $Q_k$, which includes both $E_k$ and its
$3^{d}-1$ neighboring elements. Define $H_{Q_k}[\rho]$ to be the
restriction of $H[\rho]$ to $Q_k$ with periodic boundary condition
and with potential given by the restriction of $V_{\eff}[\rho]$ to
$Q_{k}$.  $H_{Q_k}[\rho]$ is then discretized and diagonalized with
uniform basis functions such as planewaves. We denote the
corresponding eigenvalues and eigenfunctions by $\{
\lambda_{k,j}\}_{j\ge 1}$ and $\{ \varphi_{k,j}(x)\}_{j\ge 1}$,
respectively, starting from the lowest eigenvalue. One then restricts
the first $J_k$ functions of $\{ \varphi_{k, j}(x)\}_{j \ge 1}$ to
$E_k$, where $J_k$ is set to be proportional to the number of
electrons inside the extended element $Q_k$ (see the numerical
examples for specific choice of $J_k$). In addition, we define for
each $E_k$
\begin{equation} \label{eq:lambdacut}
  \lambda_k^{c} = \lambda_{k,J_k},
\end{equation}
\ie the largest selected eigenvalue in $E_{k}$ which shall be used later.
Applying the Gram-Schmidt procedure to $\{ \varphi_{k, j}(x)\}_{1 \le
  j \le J_k}$ then gives rise to a set of orthonormal functions
\begin{equation} \label{eq:ukj}
  \{u_{k, j}(x)\}_{1\le j \le J_k}
\end{equation}
for each $E_k$. The union of such functions over all elements
$\{u_{k,j}(x)\}_{1\le k \le M, 1\le j \le J_k}$ gives the set of
adaptive local basis functions (ALBs).

For a given system, the partition of $E_{k}$ is kept to be the same
even with changing atomic configurations as in the case of structure
optimization and molecular dynamics. Dangling bonds may form when
atoms are present on the surface of the extended elements, but we
emphasize that these dangling bonds are not needed to be passivated by
introducing auxiliary atoms near the surface of the extended
elements~\cite{ZhaoMezaWang2008}.  This is because the potential is
not obtained self-consistently within $Q_{k}$, but instead from the
restriction of the screened potential in the global domain $\Omega$ to
$Q_{k}$ in each SCF iteration, which mutes the catastrophic damage of
the dangling bonds.  The oscillation in the basis functions caused by
the discontinuity of the potential at the surface of the $Q_k$ (called
Gibbs phenomenon) still exists, but it damps exponentially away from
the surface of $Q_k$ and has controlled effect in $E_k$.  Using
disordered Na and Si as examples, we demonstrated that ALB can achieve
meV accuracy per atom using $4\sim 40$ basis functions per
atom~\cite{LinLuYingE2012}.

\section{Element orbitals}\label{sec:EO}

The high accuracy of ALBs indicates that the span of $\{u_{k,
  j}\}_{1\le k\le M, 1\le j\le J_k }$ approximately contains the span
of the Kohn-Sham orbitals $\{\psi_i\}_{1\le i \le
  \tilde{N}}$. However, we found that the number of basis functions
per atom may vary significantly with respect to the dimensionality $d$
of the system, which has not been seen reported in the literature
using traditional contracted basis set to the extent of our knowledge.

The dimension dependence of ALBs can be intuitively understood as
follows, motivated from the success of the contracted basis set
such as atomic orbitals. Consider the case where an atom is positioned
at the center of element $E_k$ and assume for simplicity that each of
its atomic orbital overlaps only with the neighboring elements (i.e.,
those inside the extended element $Q_k$). In order to include one such
atomic orbital denoted by $\eta(x)$ in the span of $\{u_{k,j}(x)\}_{1\le k \le M,
  1\le j \le J_k}$, each neighboring element $E_{k'}$ in $Q_k$ should
allocate one of its ALBs for representing the restriction of $\eta(x)$
in $E_{k'}$. This implies that $N^b$, the total number of ALBs, should
roughly be equal to $3^d \tilde{N}$, which becomes increasingly redundant
with respect to the dimension $d$. In fact, this is close to what has
been observed in the numerical experiments~\cite{LinLuYingE2012}.

In order to avoid this redundancy and motivated by the construction of
atomic orbitals, we propose to build a new basis set by piecing the
ALBs in neighboring elements $\{E_{k'}\}$ in $Q_k$ to construct functions
that are qualitatively close to the atomic orbitals. To distinguish
them from the pre-fitted atomic orbitals, we name these functions {\em
  element orbitals} (EOs). In order to achieve this, one is faced
mainly with three issues. First, the ALBs are always discontinuous
across the element boundaries, while qualitatively the EOs should be a
{\em continuous} function since the atomic orbitals are
continuous. Second, when one pieces back the ALBs to obtain the EOs,
it is essential that the resulting functions have {\em
  low-energy}. Finally, one needs to make sure that the EOs of $E_k$
should be {\em localized} at $E_k$ in order to avoid degeneracy.

A two-step procedure is proposed to address these three issues.  In
the first step, we construct, for each element $E_k$, a set of {\em
  candidate functions} that take care of the first two issues. Then in
the second step, we identify the element orbitals by localizing the
candidate functions. More specifically, the method proceeds as
follows.


Let us fix an element $E_k$. First, since each ALB is only supported
in its associated element and equal to zero outside, we seek for a set
of {\em candidate functions} for element $E_k$ that are linear
combinations of the ALBs of both $E_k$ and its $3^{d}-1$ neighbors
(Fig.~\ref{fig:cartoon}). Denoting by $\I$ the index of all the ALBs, and
by $\I_k \subset \I$ the index set of ALBs supported in $Q_k$, we
define a local Hamiltonian
\[
\H_k = \H(\I_k,\I_k),
\]
i.e., the restriction of $\H$ to the index set $\I_k$. Following the
intuition that the atomic orbitals should only be affected by the
local environment of $E_k$, it is reasonable to assume that the low
eigenfunctions of $\H_k$ serve as good candidate functions.
Computationally, we diagonalize $\H_k$ by
\begin{equation} \label{eqn:candidate}
  \H_k \M_k = \M_k \mathsf{\Delta}_k, 
\end{equation}
where the diagonal of $\mathsf{\Delta}_k$ contains all the eigenvalues
bounded from above by the cut-off energy $\lambda_k^c$ given by
\eqref{eq:lambdacut} and the columns of $\M_k$ contains the
corresponding eigenfunctions.  The matrix $\M_k$ is called the {\em
  merging matrix} for element $E_k$. We argue that this step addresses
the continuity and low-energy issues of the element orbitals since the
eigenfunctions \eqref{eqn:candidate} are qualitatively smooth due to
the cross-element penalty term of the DG formulation. Choosing the
eigenfunctions below $\lambda_k^c$ ensures that the candidate functions
have low-energy.

Second, we localize these candidate functions to be centered at $E_k$
using a penalizing weight function $w_k(x)$ defined for $x\in Q_k$.
$w_k(x)$ is only nonzero in the extended element $Q_k$ outside a
certain distance, called the {\em localization radius}, from the
boundary of $E_{k}$ (light gray area in Fig.~\ref{fig:cartoon}).  For
simplicity we choose $w_k(x)=1$ in the penalty area and $0$ otherwise.
More sophisticated weighting function as developed for linear scaling
methods~\cite{Garc'ia-CerveraLuXuanEtAl2009} and confining potentials
as developed for atomic orbitals~\cite{Junquera:01} can be used and
optimized for EOs in the future work.  A weighting matrix $\W_k$ for
the adaptive basis functions in the index set $\I_k$ is defined in the
extended element $Q_k$ by
\[
\W_k(k',j';k'',j'') = \average{u_{k',j'}, w_k \cdot  u_{k'',j''}}_{\mc{T}}.
\]
In order to localize the candidate functions, we solve a second
eigenvalue problem
\[
( \M_k^t \W_k \M_k) \L_k = \M_k^t \M_k \L_k\mathsf{\Gamma}_k = \L_k \mathsf{\Gamma}_k,
\]
where $\M_k^t \M_k = \I$ since $M_k$ are orthogonal from
\eqref{eqn:candidate}. The columns of $\L_k$ and the diagonal of
$\mathsf{\Gamma}_k$ consist of the first $N_k^o$ eigenfunctions and
eigenvalues, respectively. Here $N_k^o$ is the number of element
orbitals (EOs) of $E_k$. As will be shown later in the numerical
results, a small number of EOs per atom already achieve high accuracy
in the total energy calculation. We call the matrix $\L_k$ the {\em
  localization matrix}, and the product $\M_k \L_k$ gives the
coefficients of the EOs in $E_k$ in terms of the ALBs indexed by
$\I_k$. In order to present these EOs in terms of the whole adaptive
basis set, we introduce an $|\I|\times |\I_k|$ {\em selection matrix}
$\S_k$ such that $\S_k(\I_k,\I_k)$ is equal to the identity and all
zero otherwise. By defining the $N^b \times N_k^o$ coefficient matrix
$\C_{k}=\S_{k}\M_k\L_k$, we can construct the element orbitals
associated with $E_k$ by
\begin{equation}
  \phi_{k,l}(x) = \sum_{k',j'} u_{k',j'}(x) (\C_k)_{k'j';l},\quad
  l=1,\ldots,N^{o}_{k}.
  \label{}
\end{equation}
Note that, since these functions are localized in $Q_k$ by
construction, the index $k'$ only runs through the elements insides
$Q_k$. Finally, the coefficient matrix
\[
\C = \left(  \C_{1},\ldots,\C_M \right)
\]
gives the whole set of coefficients of the $N^o = \sum_{k=1}^M N_k^o$
element orbitals based on adaptive local basis functions. 
Once the element orbitals are identified, we solve an $N^o \times N^o$
generalized eigenvalue problem
\begin{equation}
(\C^t \H \C ) \V = (\C^t \C) \V \mathsf{\Lambda},
  \label{eqn:GenEig}
\end{equation}
where the diagonal of $\mathsf{\Lambda}$ gives the Kohn-Sham
eigenvalues $\{\lambda_i\}_{1\le i \le \tilde{N}}$ and the columns of
$\V$ provide the coefficients of Kohn-Sham orbitals in terms of
EOs. From $\{\lambda_i\}_{1\le i \le \tilde{N}}$, one can calculate
the chemical potential $\mu$ and the occupation number $\{f_i\}_{1\le
  i \le \tilde{N}}$. Finally, by introducing the Gram matrix
\[
\G = \C\V \cdot \text{diag}(f_i) \cdot (\C\V)^t,
\]
we can write $\rho(x)$ as
\begin{equation}
\rho(x) = \sum\nolimits_{j',j'} u_{k(x),j'}(x) \cdot \G(k(x),j'; k(x),j) \cdot u_{k(x),j}(x),
  \label{eqn:Den}
\end{equation}
where $k(x)$ indexes the element that contains $x$. Notice that one only
needs the knowledge of the diagonal blocks of the Gram matrix $\G$ to
construct the electron density. This allows us to use the recently
developed pole expansion and selected inversion type fast
algorithms~\cite{LinLuCarE2009,LinLuYingE2009,LinLuYingEtAl2009,LinYangLuEtAl2011,LinYangMezaEtAl2010,LinChenYangHe}
to reduce the asymptotic scaling for solving the generalized eigenvalue
problem~\eqref{eqn:GenEig} from cubic scaling to at most quadratic
scaling for 3D bulk systems.

\section{Parallel Implementation}\label{sec:parallel} 

Our algorithm is implemented fully in parallel for message-passing
environment, based on the implementation details presented in
Ref.~\onlinecite{LinLuYingE2012}. Here we summarize the key components
of the parallel implementation.

The global domain is discretized with a uniform Cartesian grid with a
spacing fine enough to capture the local oscillations of the Kohn-Sham
orbitals and the electron density.  Rather than using the dual grid
approach with one set of grid for representing the Kohn-Sham
wavefunctions, and another set of denser grid for representing the
electron density, we only use one set of Cartesian grid for both the
Kohn-Sham wavefunctions and the electron density for simplicity of the
implementation. The grid inside an element $E_k$ is a
three-dimensional Cartesian Legendre-Gauss-Lobatto (LGL) grid in order
to accurately carry out the operations of the basis functions such as
numerical integration.  The ALBs are first
represented in a planewave basis set in each extended element $Q_k$
solved by LOBPCG algorithm~\cite{Knyazev:01} with a
preconditioner~\cite{TeterPayneAllan1989}, and are interpolated to
each element $E_{k}$ and orthogonalized. The eigenvalue problems
involved in constructing the EOs are performed by LAPACK
subroutine \textsf{dsyevd}.

To simplify the discussion of the parallel implementation, we assume
that the number of processors is equal to the number of elements. It is
then convenient to index the processors $\{P_k\}$ with the same index
$k$ used for the elements. In the more general setting where the number
of elements is larger than the number of processors, each processor
takes a couple of elements and the following discussion will apply with
only minor modification. Each processor $P_k$ locally generates and stores the
ALBs $\{u_{k,j}(x)\}$ for $j=1,2,\ldots,J_k$
and the coefficients for the EOs $\{\C_{k;j,l}\}$ for
$j=1,2,\ldots,J_k$ and $l=1,2,\ldots,N_{k}^{o}$.  The EOs 
$\{\phi_{k,l}(x)\}$ are not explicitly formed in the real space.  We
further partition the non-local pseudopotentials $\{b_{\ell}(x)\}$ by
assigning $b_{\ell}(x)$ to the processor $P_k$ if and only if the atom
associated to $b_{\ell}(x)$ is located in the element $E_k$. 

Since the matrices $\C$ and $\H$ are sparse, the Hamiltonian matrix
$\C^t \H \C$ and the mass matrix $\C^{t} \C$ in \eqref{eqn:GenEig} are
also sparse matrices.  However, these matrices are treated as dense
matrices in our implementation for simplicity.  The parallel
matrix-matrix multiplication for constructing $\C^t \H \C$ and $\C^{t}
\C$ are performed using PBLAS subroutine \textsf{pdgemm}, and the
generalized eigenvalue problem~\eqref{eqn:GenEig} is solved by
converting it to a standard eigenvalue problem using
ScaLAPACK~\cite{ScaLAPACK}
subroutine \textsf{pdpotrf} and \textsf{pdsygst}, and the standard
eigenvalue problem is solved by ScaLAPACK subroutine \textsf{pdsyevd}.

In our implementation, the matrices $\H$ and $\C$ are constructed
locally according to the element indices. However, the ScaLAPACK
routines that operate on $\H$ and $\C$ require them to be stored in
the two dimensional block cyclic pattern. In order to support these
two types of data storage, we have implemented a rather general
communication framework that only requires the programmer to specify
the desired non-local data.  This framework then automatically fetches
the data from the processors that store them locally. The actual
communication is mostly done using asynchronous communication routines
\textsf{MPI\_Isend} and \textsf{MPI\_Irecv}.

\section{Numerical results}\label{sec:num}

The new method is implemented with Hartwigsen-Goedecker-Hutter (HGH)
pseudopotential~\cite{HartwigsenGoedeckerHutter1998}, with the local
and nonlocal pseudopotential implemented fully in the real
space~\cite{PaskSterne2005}.  Finite temperature formulation of the
Kohn-Sham density functional theory~\cite{Mermin1965} is used, and the
temperature is set to be $2000$K only for the purpose of accelerating
the convergence of SCF iteration.  Since finite temperature is used,
the accuracy is quantified by the error of the total free
energy~\cite{AlaviKohanoffParrinelloEtAl1994} per atom.  HGH
pseudopotential has analytic expression, which allows us to minimize
the effect of numerical interpolation and perform accurate comparison
with existing electronic structure code.  We compare our result with
ABINIT~\cite{Abinit1} which also supports HGH pseudopotential.  The
ALBs and EOs start from random initial guess, and are refined
iteratively in the SCF iteration together with the electron density.
In all the calculations, Anderson mixing~\cite{Anderson1965} with
Kerker preconditioner~\cite{Kerker1981} are used for the SCF
iteration.  Gamma point Brillouin sampling is used for simplicity.  In
Section~\ref{sec:ALB} and Section~\ref{sec:EO}, we count the number of
basis functions in terms of the number of ALBs per element and the
number of EOs per element.  In this section, we count the number of
ALBs and EOs per atom instead, in order to be consistent with
literature.  All computational experiments are performed on the Hopper
system at the National Energy Research Scientific Computing (NERSC)
center.  Each Hopper node consists of two twelve-core AMD ``MagnyCours''
2.1-GHz processors and has 32 gigabytes (GB) DDR3 1333-MHz
memory. Each core processor has 64 kilobytes (KB) L1 cache and 512KB
L2 cache. It also has access to a 6 megabytes (MB) of L3 cache shared
among 6 cores.

As mentioned earlier, the ALBs have been shown to achieve effective
dimension reduction for quasi-1D systems, but with deteriorating
performance as the dimensionality of the system
increases~\cite{LinLuYingE2012}.  Using Na as example, it has been shown
that while $4$ ALBs per atom is enough to reach $1$ meV accuracy for
quasi-1D systems, $40$ ALBs per atom is necessary to reach the same
accuracy for 3D bulk systems.  Now using a 3D bulk Na system with $432$
atoms as example, we illustrate that the number of basis functions per
atom can be effectively reduced using EO.

\begin{figure}[h]
  \begin{center}
   \includegraphics[width=0.5\textwidth]{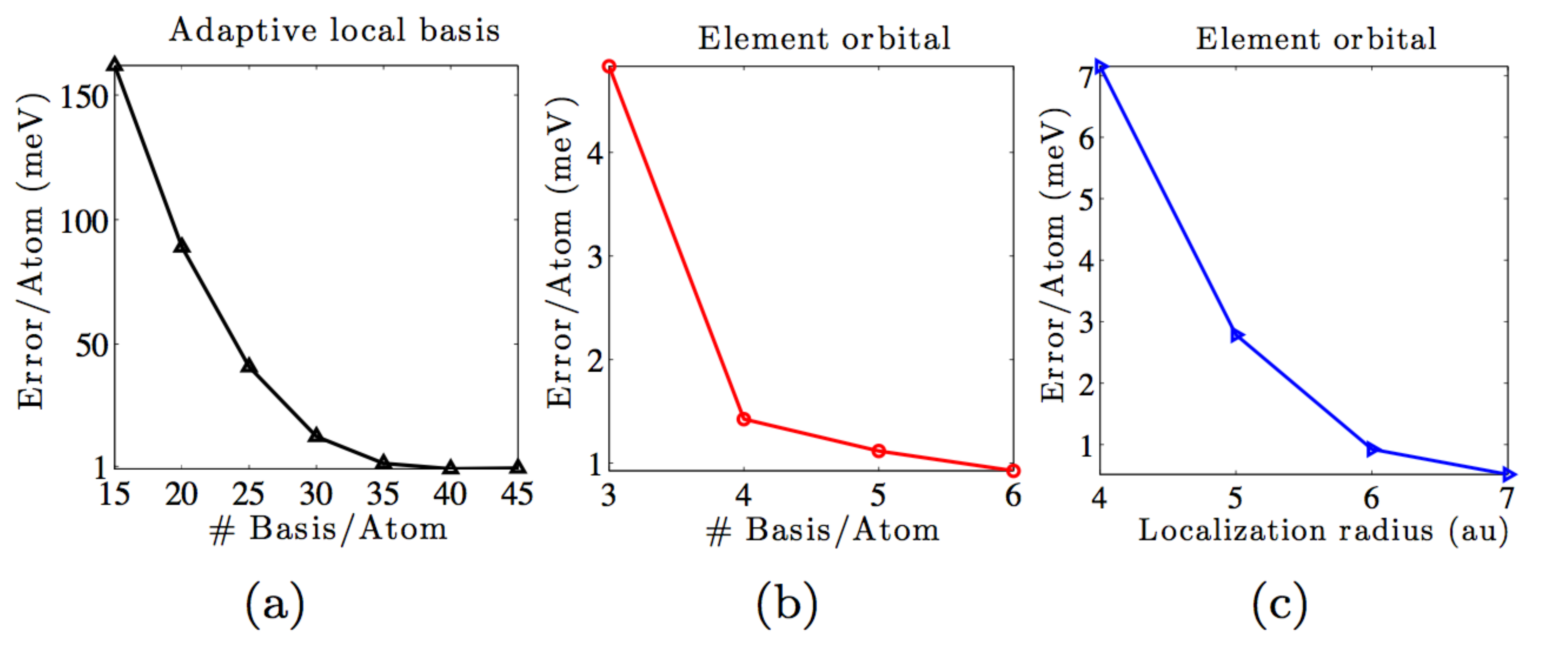} 
  \end{center}
  \caption{(color online) (a) Convergence of adaptive local basis
  functions (ALB) for a 3D bulk Na system with $432$ atoms.  (b) Convergence
  of element orbitals (EO) for the same Na system with fixed number of
  ALBs.  (c) Convergence in terms of the localization radius for the
  same Na system with fixed number of ALBs and fixed number of EOs.}
  \label{fig:NaConv}
\end{figure}

The supercell for Na is simple cubic and the length of the supercell
along each dimension is $45.6$ \au.  A random perturbation with
standard deviation $0.2$ \au\ is applied to each atom in the supercell
to eliminate the translational invariance of the system.  The
supercell is partitioned into $6\times 6\times 6$ elements, with the
length of each dimension of each element being $7.6$ \au.  The length
of each dimension of each extended element is $22.8$ \au\ which is $3$
times larger than that of the element.  The penalty parameter $\alpha$
in \eqref{eqn:KSDG} is set to be $100$. The supercell is discretized with
a uniform mesh of dimension $120\times 120\times 120$ in the real
space.  This mesh is used for representing both the electron density
and the Kohn-Sham orbitals, which corresponds to a planewave cutoff
of $68$ Ry in the Fourier space.  ABINIT uses a dual grid for
representing the Kohn-Sham wavefunctions and the electron density.
The planewave cutoff for wavefunctions used in ABINIT is $20$ Ry. This
corresponds to a planewave cutoff for the electron density at $80$ Ry,
with a uniform mesh of dimension $135\times 144\times 144$ in the real
space.  The different numbers of grid points along each dimension come
from the automatic grid adjustment in ABINIT.  We remark that the grid
size is chosen to be larger than the typical setup in electronic
calculation for Na to make sure that the error introduced by the grid
size is small compared to that introduced by using ALBs and EOs.
Inside each element a Legendre-Gauss-Lobatto (LGL) grid of dimension
$30\times 30\times 30$ is used for numerical integration in the
assembly process of the discretized Hamiltonian matrix $\H$.  The
error of the total free energy per atom only using ALBs is shown in
Fig.~\ref{fig:NaConv} (a).  The error systematically decreases with
the increase of the number of ALBs. When the number of ALBs exceeds
$35$, the error of the total free energy per atom is less than $1$
meV.

Element orbitals (EOs) provide further dimension reduction compared to
ALBs.  Fig.~\ref{fig:NaConv} (b) shows the difference of the free
energy per atom calculated from EOs and that from ABINIT.  We
construct EOs from as many as $42$ ALBs per atom, following the
criterion \eqref{eq:lambdacut} for the choice of the candidate
functions and using a localization radius of $6.0$ \au.  Compared to a
converged ALB calculation, the error using only $3$ EOs per atom is
already within $5$ meV per atom.  When $6$ EOs are used, the total
free energy calculated is essentially the same as that using $42$
ALBs, and the error compared to ABINIT is less than $1$ meV per atom.
Fig.~\ref{fig:NaConv} (b) indicates that the EOs are
indeed effective for reducing the number of basis functions per atom
for 3D bulk systems.

Compared to ALB, the EO approach introduces an additional
parameter which is the localization radius.  Fig.~\ref{fig:NaConv} (c)
shows the error of the total free energy per atom using $42$ ALBs per
atom, and $6$ EOs per atom but with different localization radius.
When the localization radius is $4.0$ \au\ which is $53\%$ the length
of an element, the error of the total energy per atom is $7$ meV.
Moderate choice of the localization radius of $6.0$ \au\ ($69\%$ of the
length of an element) yields accuracy around $1$ meV per atom.
Fig.~\ref{fig:NaConv} (c) shows that our method is stable even for a
large localization radius $7.0$ \au\ ($92\%$ of the length of an
element), and the error is even smaller and is below $1$ meV per atom.
We also remark that if the localization radius is further increased,
the EOs are no longer localized around the element, but
become fully extended in the extended element.  This can lead to an
unstable scheme with large error.  Fig.~\ref{fig:NaConv} (c) shows
that the accuracy of the EO is not very sensitive to the
choice of localization radius.

\begin{figure}[h]
  \begin{center}
    {\includegraphics[width=0.5\textwidth]{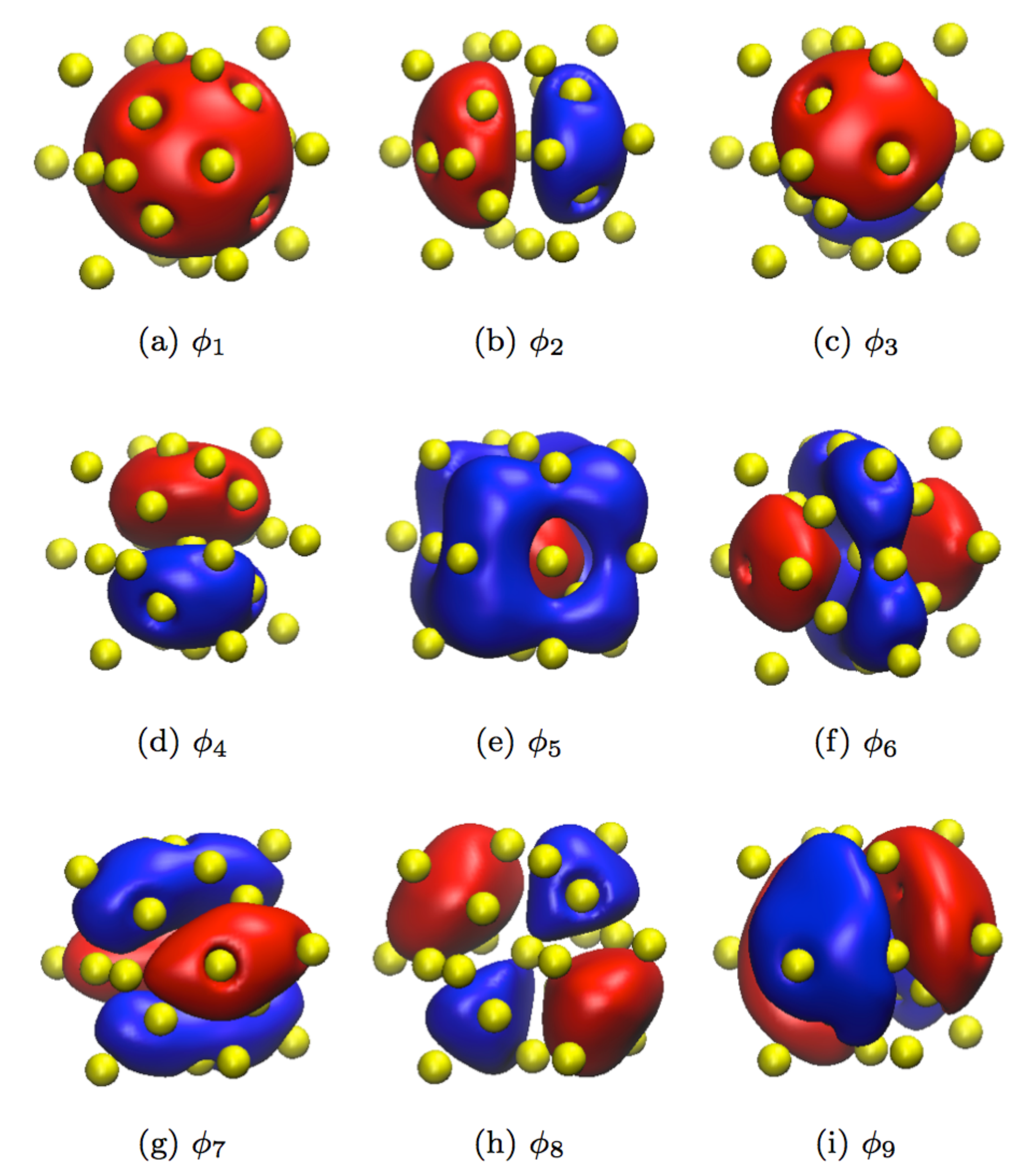}}
  \end{center}
  \caption{The isosurface of the first $9$ element orbitals belonging to
  the same extended element, for a 3D bulk Na system with $432$ atoms.
  The $27$ Na atoms nearest to the element orbitals within
  a sphere of radius $6.0$ \au\ are plotted as gold balls.  The positive
  and negative part of the element orbitals are represented by red and
  blue color, respectively.}
  \label{fig:basis}
\end{figure}

EOs can resemble atomic orbitals but with local
modifications reflecting the environmental effect, despite the fact
that they are constructed in the extended elements with rectangular
domain.  Using the same Na system as example, we show in
Fig.~\ref{fig:basis} the isosurface of the first $9$ element orbitals
($\phi_1$ to $\phi_{9}$) belonging to the same extended element, with
the red and blue color indicating the positive and negative part of
the EOs, respectively.  $27$ atoms nearest to these
EOs within a sphere of radius $6.0$ \au\ are also plotted
in Fig.~\ref{fig:basis} as gold balls.  We see that $\phi_{1}$ mimics
$s$ orbital, $\phi_{2}$-$\phi_{4}$ mimic $p$-orbitals, and
$\phi_{5}$-$\phi_{9}$ mimic $d$-orbitals.  Both the general shape and
the multiplicity of the element orbitals agree well with the physical
intuition.  We also find that hybridization of the $s,p,d$ orbitals
naturally appears in the EOs, reflecting the effect of
the environment.  For example, the isosurface of $\phi_{1}$ exhibits
``holes'' around atoms. These holes are not described in the spherical
symmetric $s$ atomic orbital, but can only be reflected in orbitals of
higher angular momentum such as $d$ orbitals.  Therefore, 
EOs are natural generalization of atom-centered orbitals, with
both the atomic and environmental effect taken into account
simultaneously.

\begin{figure}[h]
  \begin{center}
    {\includegraphics[width=0.5\textwidth]{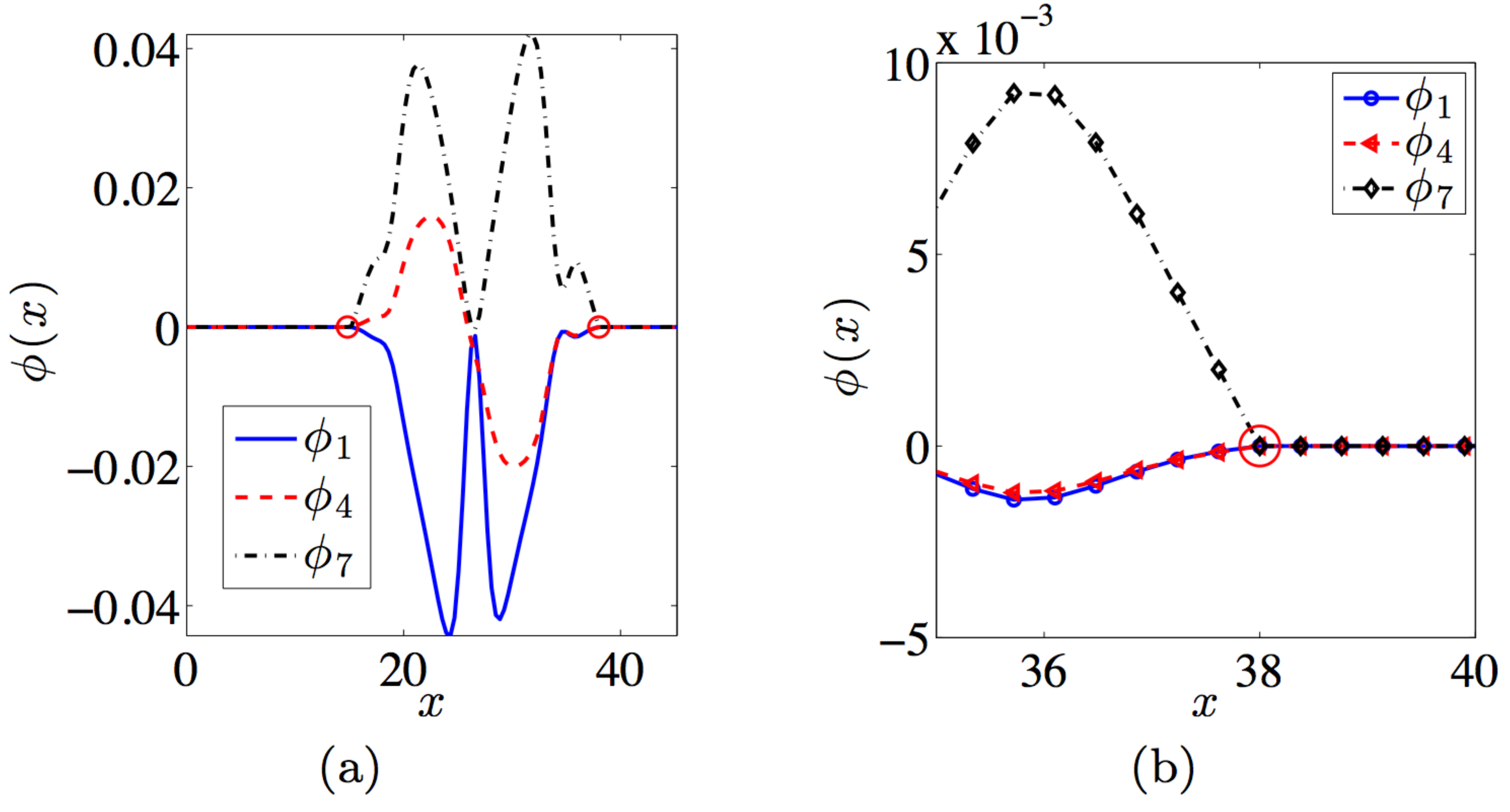}}
  \end{center}
  \caption{(color online) (a) The value of the element orbitals
  $\phi_{1}$ (blue solid line),$\phi_{4}$ (red dashed line), and
  $\phi_{7}$ (black dot dashed line) along one $[100]$ direction of a 3D bulk
  Na system with $432$ atoms. The two red circles indicate the boundary
  of the extended element.  (b) Zoom-in of (a) to the region near the
  boundary of the extended element. The same set of element orbitals
  $\phi_{1}$ (blue solid line with circles), $\phi_{4}$ (red dashed line
  with triangles) and $\phi_{7}$ (black dot dashed line with diamonds)
  are shown, with the symbols indicating the position of the numerical
  grids. The red circle indicates the boundary of the extended element.}
  \label{fig:basis1D}
\end{figure}

EOs are localized in the extended elements.  Since
each candidate function is not continuous across the boundary of the
extended element, EOs are still discontinuous across
the boundary of the extended element.  Nonetheless, the EOs are ``qualitatively continuous'' at the boundary of the extended
elements.  Fig.~\ref{fig:basis1D} (a) shows the behavior of
$\phi_{1},\phi_{4},\phi_{7}$ for the Na system along one $[100]$
direction, with the zoom-in near the boundary of the extended element
shown in Fig.~\ref{fig:basis1D} (b).  EOs are very
close to a continuous function especially for $\phi_{1}$ and
$\phi_{4}$ with lower angular momentum.  The value of EOs of higher
angular momentum such as $\phi_{7}$ at the grid point closest to the
boundary of the extended element is within $10^{-3}$.

\begin{figure}[h]
  \begin{center}
    {\includegraphics[width=0.5\textwidth]{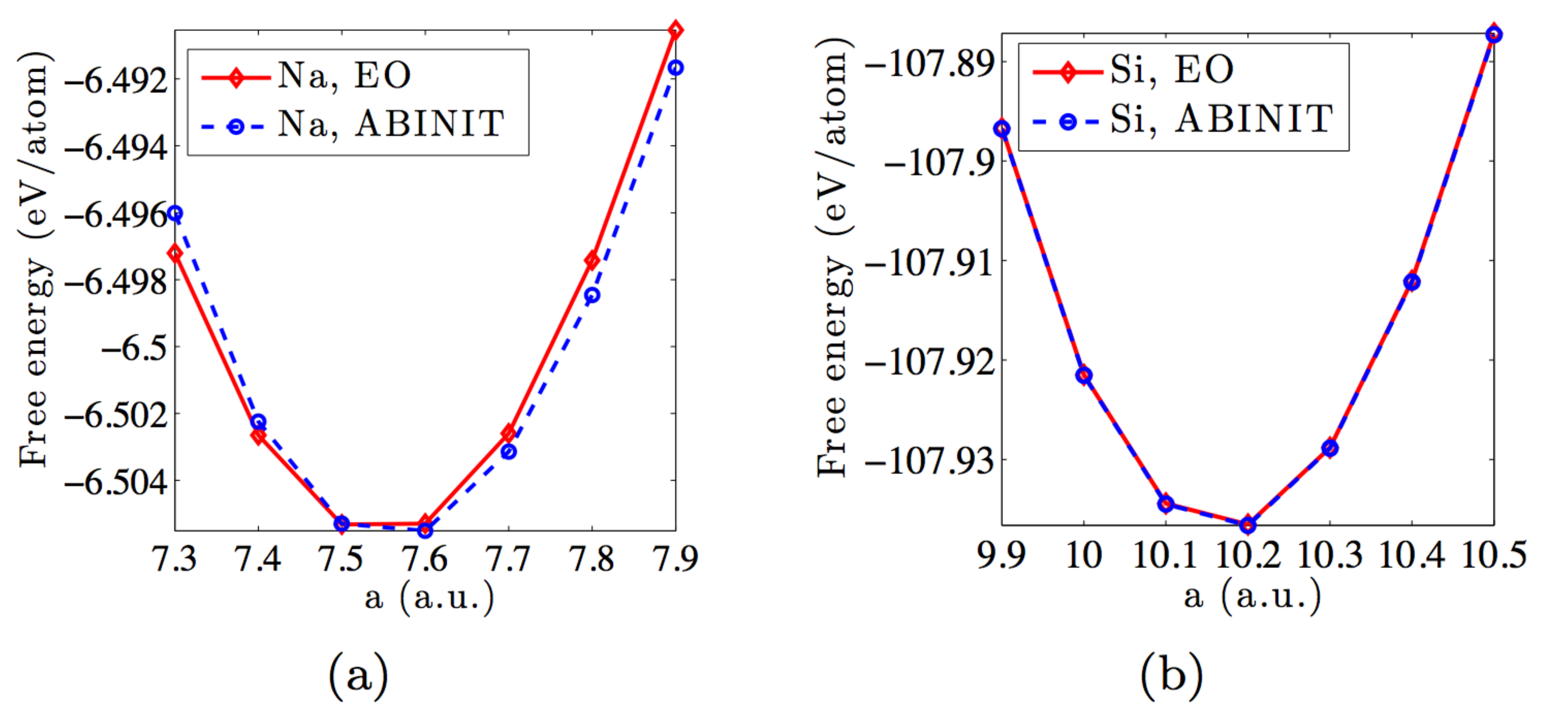}}
  \end{center}
  \caption{(color online) The total free energy per atom for 3D bulk Na
  system with $432$ atoms (a) and 3D bulk Si system with $512$ atoms
  (b), with different lattice constants calculated from ABINIT and from
  element orbitals.}
  \label{fig:latconst}
\end{figure}

EOs can be used for calculating the relative energies
of different atomic configurations.  Fig.~\ref{fig:latconst} (a) shows
the total free energy per atom for a crystal of Na consisting of
$6\times 6\times 6 = 216$ unit cells with $432$ atoms. Each unit cell
is body centered cubic with $2$ Na atoms.  The lattice constant ranges
from $7.3$ \au\ to $7.9$ \au. The size of each element is equal to that
of one unit cell.  $4$ EOs per atom are constructed from $42$ ALBs per
atom and are used for calculating the total free energy.  The planewave
cutoff
for Kohn-Sham wavefunctions in ABINIT is $20$ Ry. The difference of
the total energy per atom is less than $2$ meV across all the lattice
constants.  Similar result can be obtained for Si.  The supercell for
Si contains $4\times 4\times 4=64$ unit cells with $512$ atoms in
total.  Each unit cell is diamond cubic with $8$ Si atoms.
Fig.~\ref{fig:latconst} (b) reports the total free energy per atom for
lattice constants from $9.9$ \au\ to $10.5$ \au.  Each element only
covers $\frac23\times \frac23\times \frac23$ unit cells.  We remark
that elements occupying a fraction of the unit cell are allowed, which
is important especially when EOs are applied to systems
with defects and disorderedness.  The planewave cutoff for Kohn-Sham
wavefunctions in ABINIT is set to be $120$ Ry to achieve the high
accuracy as benchmark solution.  The localization radius is also $6.0$
\au.  Starting from $50$ ALBs per atom, $10$ EOs per atom are
computed.  The difference of the total free energy per atom is less
than $1$ meV for all lattice constants.

EOs are also effective for calculating the total energy of
systems with defects.  For a crystal Na system with $432$ atoms and the
length of each dimension of the supercell being $45.6$ \au, the total free
energy evaluated using ABINIT is $-103.27947$ \au. Using the same setup
as done in the crystal system with $4$ EOs per atom, the total free
energy evaluated using EO is $-103.27588$ \au.  The difference is as
small as $0.22$ meV per atom.  Since our implementation takes the
spin-unpolarized form, we consider a system with two vacancies by
removing $2$ Na atoms belonging to one unit cell from the supercell.
All the parameters are the same as those for the calculation of the
crystal system.  The total free energy evaluated using ABINIT is
$-102.76957$ \au, and the total free energy evalauted using $4$ EOs per
atom is $-102.76637$ \au, with the difference being $0.20$ meV per atom.
The error for both the crystal and the defect system is 
less than $1$ meV per atom.  We also estimate the formation
energy of $M$ neutral vacancies by 
\begin{equation}
 \Delta E(M)=E^{d}_{N-M}-E^{0}_{N} \frac{N-M}{M},
  \label{eqn:formation}
\end{equation}
with $E^{0}_{N}$ being the free energy for the crystal system with $N$
atoms, and $E^{d}_{N-M}$ being the free energy for the same system but
with $M$ atoms removed.  Atomic relaxation is not taken into account at
this stage.  Using \eqref{eqn:formation}, the formation energy
calculated from ABINIT is $0.864$ eV, and that calculated from EO is
$0.854$ eV. The difference of the formation energy is $0.010$ eV, and the
relative error of the formation energy is $1.2\%$.

The calculation of the defect formation energy for Si is as follows.
For a crystal Si system with $512$ atoms and the length of each
dimension of the supercell being $40.4$ \au, the total free energy
evaluated using ABINIT is $-2030.85824$ \au, and the total free energy
evaluated using $10$ EOs per atom is $-2030.85691$ \au.  The difference
is as small as $0.07$ meV per atom.  A defect system is constructed by
removing one Si atom, and all the parameters are the same as those for
the crystal calculation.  The total free energy evaluated using ABINIT is
$-2026.76478$ \au, and the total free energy evaluated using $10$ EOs per
atom is $2026.75974$ \au, with the difference being $0.27$ meV per atom.
The error for both the crystal system and that for the
defect system is less than $1$ meV per atom.   The formation energy
calculated from ABINIT is $3.454$ eV, and that calculated from EO is
$3.555$ eV. The difference of the formation energy is $0.101$ eV, and
the relative error of the formation energy is $2.9\%$.  


\begin{figure}[h]
  \begin{center}
    {\includegraphics[width=0.5\textwidth]{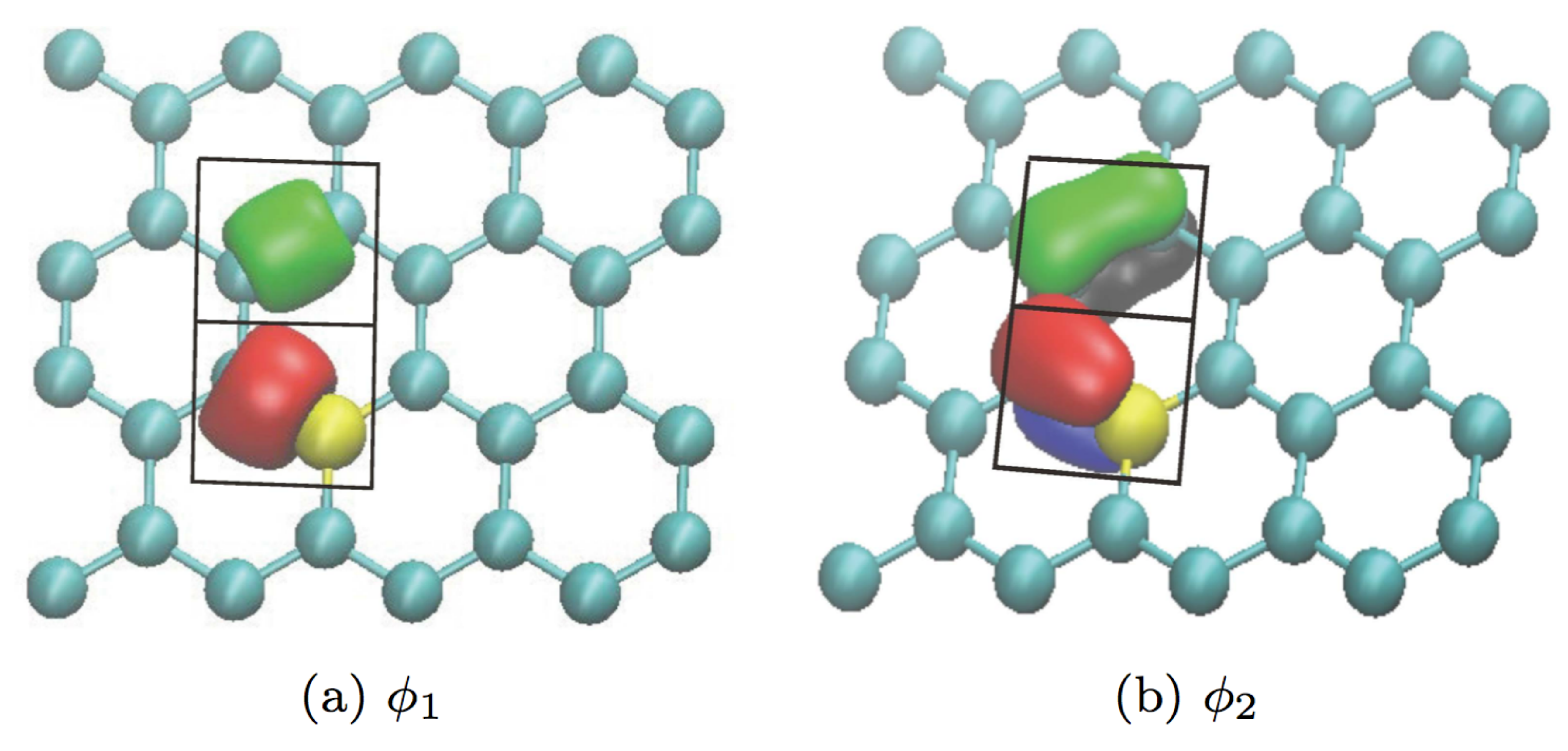}}
  \end{center}
  \caption{Graphene sheet consisting of $32$ C atoms (cyan balls) with $1$ C atom
  substituted by a Si atom (gold ball). Each black box represents an
  element.  (a) The first element
  orbital $\phi_{1}$ (green) for the upper element with $2$ C atoms, and
  the first element orbital $\phi_{1}$ (red) for the lower element with
  $1$ C atom and $1$ Si atom.  (b) The second element orbital
  $\phi_{2}$ (green for the positive part and black for the negative
  part) for the upper element with $2$ C atoms, and the second element
  orbital $\phi_{2}$ (red for the positive part and blue for the negative
  part) for the lower element with $1$ C atom and $1$ Si atom.} 
  \label{fig:GrapheneBasis}
\end{figure}

Next we study graphene sheet consisting of $32$ C atoms (cyan balls),
with $1$ C atom replaced by a Si atom (gold ball), as shown in
Fig.~\ref{fig:GrapheneBasis}.  The length of the supercell is $10.000$
\au, $16.108$ \au\ and $18.600$ \au\ for $x,y,z$ directions,
respectively.  The C and Si atoms are in the $y-z$ plane.  The
supercell consists of $4\times 4$ elements, with each element
containing $2$ atoms, and represented by one black box.  The length of
each element is therefore $10.00$ \au, $4.027$ \au\ and $4.650$ \au\ along $x,y,z$ directions, respectively.  The shape of the EOs is shown
in Fig.~\ref{fig:GrapheneBasis} (a) for the first EOs
$\phi_{1}$ belonging to $2$ different elements, and
(b) for the second EOs $\phi_{2}$ belonging to the same
$2$ elements, respectively.  We find that $\phi_{1}$ in the upper
element reflects the C-C bond and $\phi_{1}$ in the lower element
reflects the C-Si bond, respectively.  Similarly, $\phi_{2}$ reflects
the $\pi$ bonds in both the upper and the lower elements. The shape
of the EOs agree well with the physical intuition.  In
particular, the element orbitals are not centered around individual
atoms but correspond directly to chemical bonds, which are of lower
energy than individual atomic orbitals. Fig.~\ref{fig:GrapheneBasis}
shows that the EOs constructed from a complete basis set
such as planewaves provides a more flexible treatment of chemical
environment than atom centered orbitals.  The total free energy
calculated using ABINIT with a planewave cutoff at $200$ Ry is
$-180.56324$ \au.  $12$ EOs per atom contracted from $40$ ALBs per
atom with localization radius being $3.0$ \au.  The total free energy
calculated using EO is $-180.56279$ \au.  The difference in the total
free energy per atom is $0.38$ meV.

\begin{figure}[h]
  \begin{center}
    {\includegraphics[width=0.5\textwidth]{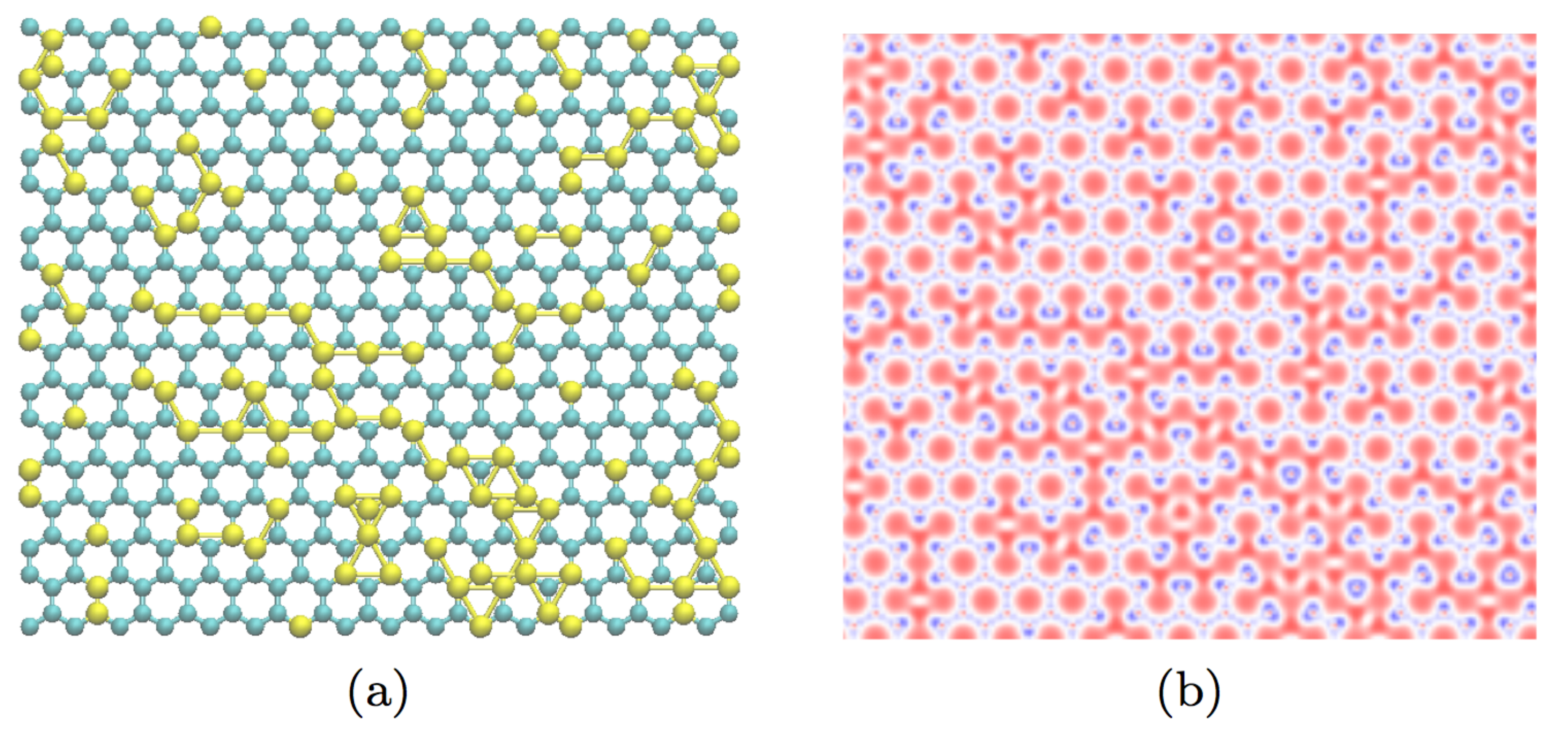}}
  \end{center}
  \caption{(a) The atomic configuration of a graphene
  sheet consisting of $512$ C atoms (cyan balls), with
  $128$ C atoms randomly selected and substituted by Si atoms (gold
  balls).  (b) The electron density across $y-z$ plane.}
  \label{fig:GrapheneDensity}
\end{figure}

A more complicated example is a graphene sheet with $512$ C atoms, and
with $128$ of the C atoms randomly selected and replaced by Si atoms.
The atomic configuration is shown in Fig.~\ref{fig:GrapheneDensity}
(a), with the C atoms represented by cyan balls and Si atoms
represented by gold balls, respectively. The atoms are
all in the $y-z$ plane, and the dimension of the supercell is $10.000$
\au, $64.432$ \au\ and $74.400$ \au\ along $x,y,z$ directions,
respectively.  The electron density in the $y-z$ plane is shown in
Fig.~\ref{fig:GrapheneDensity} (b).  The total free energy calculated
from ABINIT is $-2639.02487$ \au, and the total free energy calculated
from EO with $12$ EOs per atom for all elements is $-2639.11504$ \au.
The error of the total free energy per atom is $4.79$ meV per atom.

\begin{figure}[h]
  \begin{center}
    {\includegraphics[width=0.35\textwidth]{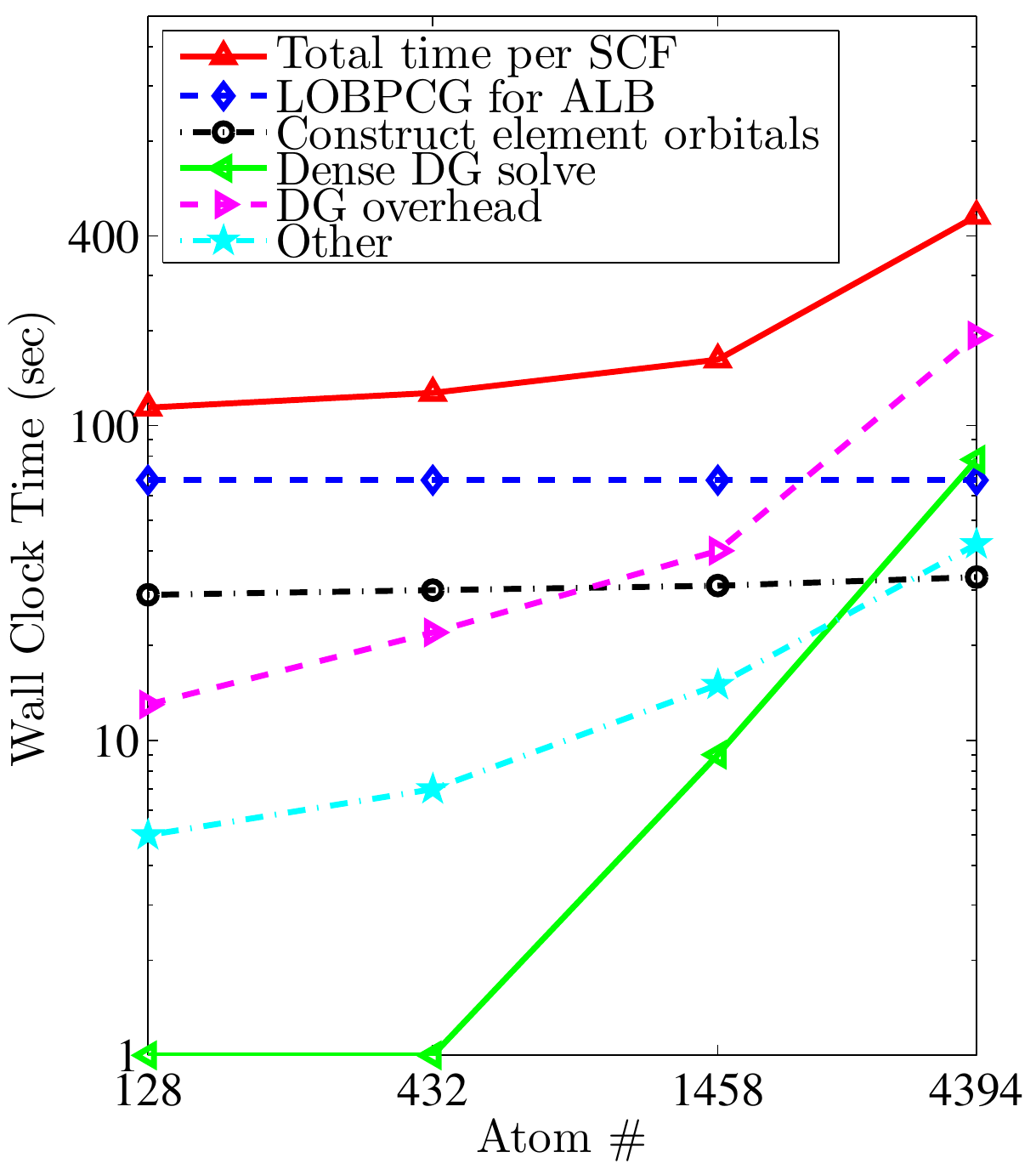}}
  \end{center}
  \caption{(color online) The total computational time per SCF iteration
  (red solid line with upward-pointing triangles) for 3D bulk Na systems
  ranging from $128$ atoms to $4394$ atoms. The breakdown of the total
  computational time includes the time for using LOBPCG to generate
  adaptive local basis functions 
  (blue dashed line with diamonds), the time for constructing the
  element orbitals from adaptive local basis functions (black dot dashed
  line with circles), the time for solving the generalized eigenvalue
  problem using dense ScaLAPACK solver (green solid line with left-pointing
  triangles), the overhead time for solving the DG problem (magenta
  dashed line with right-pointing triangles), and the rest of the time
  in a SCF iteration (cyan dot dashed line with stars).}
  \label{fig:Na3Dtime}
\end{figure}

The fact that a small number of EOs  per atom already
achieve high accuracy allows us to perform calculations for systems of
large size.  Here we study 3D bulk Na systems of various sizes,
ranging from $128$ atoms to $4394$ atoms.  The length of the supercell
along each dimension is also proportional to the system size, from
$30.4$ \au\ for $128$ atoms to $98.8$ \au\ for $4394$ atoms.  The number
of processors (computational cores) used is chosen to be proportional
to the number of atoms, with $64$ processors used for $128$ atoms, and
$2196$ processors used for $4392$ atoms. $4$ EOs per atom are
constructed from $42$ ALBs per atom for all calculations.  The total
time per SCF iteration is shown in Fig.~\ref{fig:Na3Dtime}.  We find
that even though the number of atoms increase by a factor of $34$, the
wall clock time only increases by less than $4$ times from $114$ sec
for $128$ atoms to $413$ sec for $4394$ atoms.  The small increase of
the total wall clock time is because the time for solving the
generalized eigenvalue problem~\eqref{eqn:GenEig}, which is
asymptotically the computationally dominating part, only takes less
than $100$ sec even for system as large as $4392$ atoms, thanks to the
small number of basis functions per atom allowed to be used in the
calculation.  The time for generating the ALBs using LOBPCG and the
time for constructing the EOs from the ALBs are flat for all systems,
since these steps are localized in each extended element and the
computational cost is independent of the global system size.  The
overall time for solving the generalized eigenvalue
problem~\eqref{eqn:GenEig} has not dominated the computational time
for $4392$ atoms with a Hamiltonian matrix of size $17568$.  However,
the wall clock time for this part already scales quadratically with
respect to the number of atoms.  Since the number of processors scale
linearly with respect to the system size, the overall time for solving
the generalized eigenvalue problem scales cubically with respect to
the system size, and will eventually dominate the overall running time
for systems of larger size.  The overhead of the DG calculation
involves the assembly of the DG matrix $\H$, the construction of the
Hamiltonian matrix $\C^{t}\H \C$ and the mass matrix $\C^{t}\C$ using
parallel matrix-matrix multiplication, as well as the communication
time.  As alluded to earlier, the parallel matrix-matrix
multiplication treats $\C$ and $\H$ as dense matrices in the current
implementation.  Therefore the asymptotic scaling of this part has the
same asymptotic cubic scaling as solving the generalized eigenvalue
problem.  All the rest of the computational time (classified as
``other time'' in Fig.~\ref{fig:Na3Dtime}) mainly includes constructing
the electron density using \eqref{eqn:Den} in the global domain, solving
the Kohn-Sham potential from the electron density, charge mixing as well
as the extra data communication.

\begin{figure}[h]
  \begin{center}
    {\includegraphics[width=0.5\textwidth]{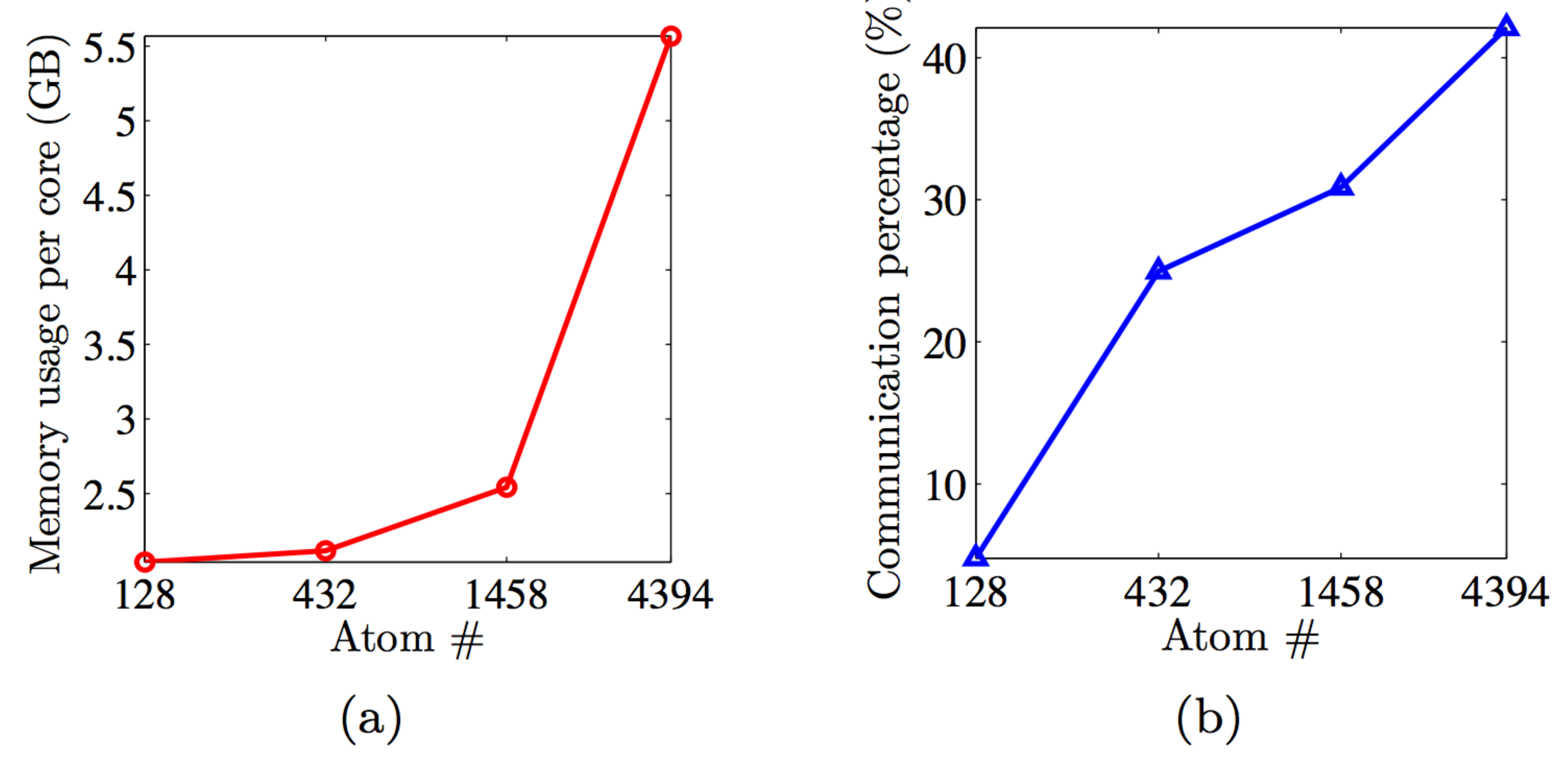}}
  \end{center}
  \caption{(color online) The memory cost per processor (a) and the
  communication percentage (b) for 3D bulk Na systems ranging from
  $128$ atoms to $4394$ atoms.} 
  \label{fig:NaMemoryCommunication}
\end{figure}

We also remark that treating the Hamiltonian matrix as dense matrices
greatly increases the memory cost and the communication volume.
Fig.~\ref{fig:NaMemoryCommunication} (a) shows the amount of memory used
per processor. When the number of atoms is $4394$, the memory used per
processor is $5.5$ GB, which becomes the bottleneck for further
increasing the system size, despite that the computational time per SCF
is still within affordable range.  The communication volume,
indicated by the percentage of the communication time within the total
computational time is shown in Fig.~\ref{fig:NaMemoryCommunication} (b).
The communication time occupies more than $40\%$ of the total time for
systems with $4394$ atoms.  Both the large memory cost and the large
communication volume is largely due to the treatment of $\C$ and $\H$ as
dense matrices, and shall be improved in the future work.

\section{Conclusion}\label{sec:conclusion}

In conclusion, we have introduced the element orbitals for
discretizing the Kohn-Sham Hamiltonian in the pseudopotential framework,
which are contracted automatically from a uniform basis set.  Comparing
with the existing contracted basis sets, element orbitals incorporate
environment information by including directly all atoms in the
neighboring elements on the fly. The implementation of element orbitals
is straightforward thanks to the rectangular partitioning of the domain.
The accuracy of element orbitals are systematically improvable and the
same procedure can be applied to systems under various conditions.  The
element orbitals are constructed by solving KSDFT locally in the real
space, and localized on each element via a localization procedure.  We
remark that the localization procedure used for constructing the element
orbitals is not grounded on the near-sightedness property as in the
linear scaling methods for insulating
systems~\cite{Goedecker1999,Kohn1996}.  Instead of finding the compact
representations for the Kohn-Sham invariant subspaces~\cite{Gygi2009},
the current work seeks for a set of compact basis functions in the real
space, while the coefficients of the basis set for representing the
Kohn-Sham orbitals can still be delocalized. As is shown by the
numerical examples, the current procedure is applicable to both
insulating and metallic systems.

Our numerical examples also indicate that treating $\C$ and $\H$ as
dense matrices can greatly increase the memory cost, the communication
volume and the computational time especially for systems of large size.
The future improvement includes treating $\C$ and $\H$ as sparse
matrices so that the construction of the Hamiltonian matrix $\C^{t}\H
\C$ and the mass matrix $\C^{t}\C$ is of linear scaling. By treating
$\C$ and $\H$ as sparse matrices, we can also incorporate the recently
developed pole expansion and selected inversion type fast
algorithms~\cite{LinLuCarE2009,LinLuYingE2009,LinLuYingEtAl2009,LinYangLuEtAl2011,LinYangMezaEtAl2010,LinChenYangHe}
to reduce the asymptotic scaling for solving the generalized eigenvalue
problem~\eqref{eqn:GenEig} from cubic scaling to at most quadratic
scaling for 3D bulk systems. We also remark that the current procedure
for constructing the orbitals from adaptive local basis
functions is still a costly procedure inside each element.  Method for
generating element orbitals directly inside the extended element is also
under our exploration.

This work is partially supported by NSF CAREER Grant 0846501 (L.~Y.),
and by the Laboratory Directed Research and Development Program of
Lawrence Berkeley National Laboratory under the U.S. Department of
Energy contract number DE-AC02-05CH11231 (L.~L.). The authors thank Jianfeng Lu
for helpful discussions, and National Energy Research Scientific
Computing Center (NERSC) for the support to perform the calculations.
L.~L. also thanks Weinan E for encouragement, and the University of Texas
at Austin for the hospitality where the idea of this paper starts.


\end{document}